# Large signal analysis of multiple quantum well transistor lasers: Investigation of imbalanced carrier and photon density distribution

**Iman Taghavi[1*], Behzad Namvar[2**], Mohammad Hosseini[2***] and Hassan Kaatuzian[2****]**

[1]*Quantum Matter Institute, University of British Columbia, 447-2355 E Mall, Vancouver, BC V6T 1Z4*

[2]*Photonics Research Lab., Electrical Engineering Dept., Amirkabir University of Technology, Hafez Ave., Tehran, 15914*

**staghavi3@ece.ubc.ca; **behzadnmvr@aut.ac.ir; ***smohammad.hsi@aut.ac.ir; ****hsnkato@aut.ac.ir*

In this paper, we present a large signal and switching analysis for the Heterojunction Bipolar Transistor Laser (TL) to reveal its optical and electrical behavior under high current injection conditions. Utilizing appropriate models for carrier transport, nonlinear optical gain and optical confinement factor (OCF), we have simulated the large signal response of the HBTL in relatively low and high modulation frequencies. Our results predict that for multiple quantum well structures at low frequencies there should not be a difference in carrier density either the photon density. However, carrier concentration can be differently distributed between subsequent wells in case of a high speed yet large signal input. This leads to increased linewidth instead as it depends on $\Delta N_{qw}$. We show the effect of different structural parameters on the switching behavior by performing a switching analysis of the Single Quantum Well (SQW) and Multiple Quantum Well (MQW) structures using computationally efficient numerical methods. A set of coupled rate equations is solved to investigate the large-signal and switching behavior of MQW-HBTL. Finally, to have a comprehensive judgment about this optoelectronic device, we introduce a relative performance factor to taking into account all the optoelectronic characteristics such as output power, ac current gain, modulation bandwidth, and base threshold current as well as turn-on time in order to design a suitable TL for Opto-Electronic Integrated Circuits (OEICs).

## 1. INTRODUCTION

The idea of light extraction from the base region of a transistor at room temperature has been initially proposed in several publications in the 80's decade of the twentieth century [1]. However, the first successful demonstration of the transistor laser (TL) idea was introduced in 2004 [2]. Since then, many researchers have focused on the TL both in theoretical and experimental aspects. TL can work as a three-port device with simultaneous electrical and optical outputs [3]. Further improvements including wavelength tunability, Multiple Quantum Well (MQW) structure, resonance-free characteristics have been proposed yet [4, 5]. While edge-emitting transistor lasers were the original type of this new family of optoelectronic devices, vertical-cavity structure of transistor laser (i.e., Vertical Cavity Transistor Laser) which was reported for the first time in 2012 [6] also attracted a considerable attention thanks to their promising features including enhanced line-width, optical power and reduced threshold current [7-9].

In the TL, as figured in Fig. 1, the usual electrical collector is accompanied by an optical collector, i.e. the QW, inserted in the base region of the HBT.

Stimulated recombination, a unique mechanism to the TL, causes "compression" in the collector I-V characteristics and optical gain decrease, which will be shortly discussed in this paper. Combined functionality of an HBT (i.e., amplification of a weak electrical signal with a high gain at high speed) and that of a diode laser (i.e., generating laser emission) is observed in the TL. In other words, a modulating base current leads to simultaneous modulated signals for both optical and electrical outputs. It raises the possibility of replacing some metal wiring between components on a circuit board or wafer chip with optical interconnections, thus providing more flexibility and capability in optoelectronic integrated circuits [10].

Owing to these interesting properties of TL, one may anticipate promising functionalities that are appropriate for telecommunication and other applications.

Aside from its capability for handling electrical and optical signals at the same time, TL has a larger optical bandwidth (BW) and a resonance-free frequency response thanks to its unique carrier transport mechanism within the base region [11].

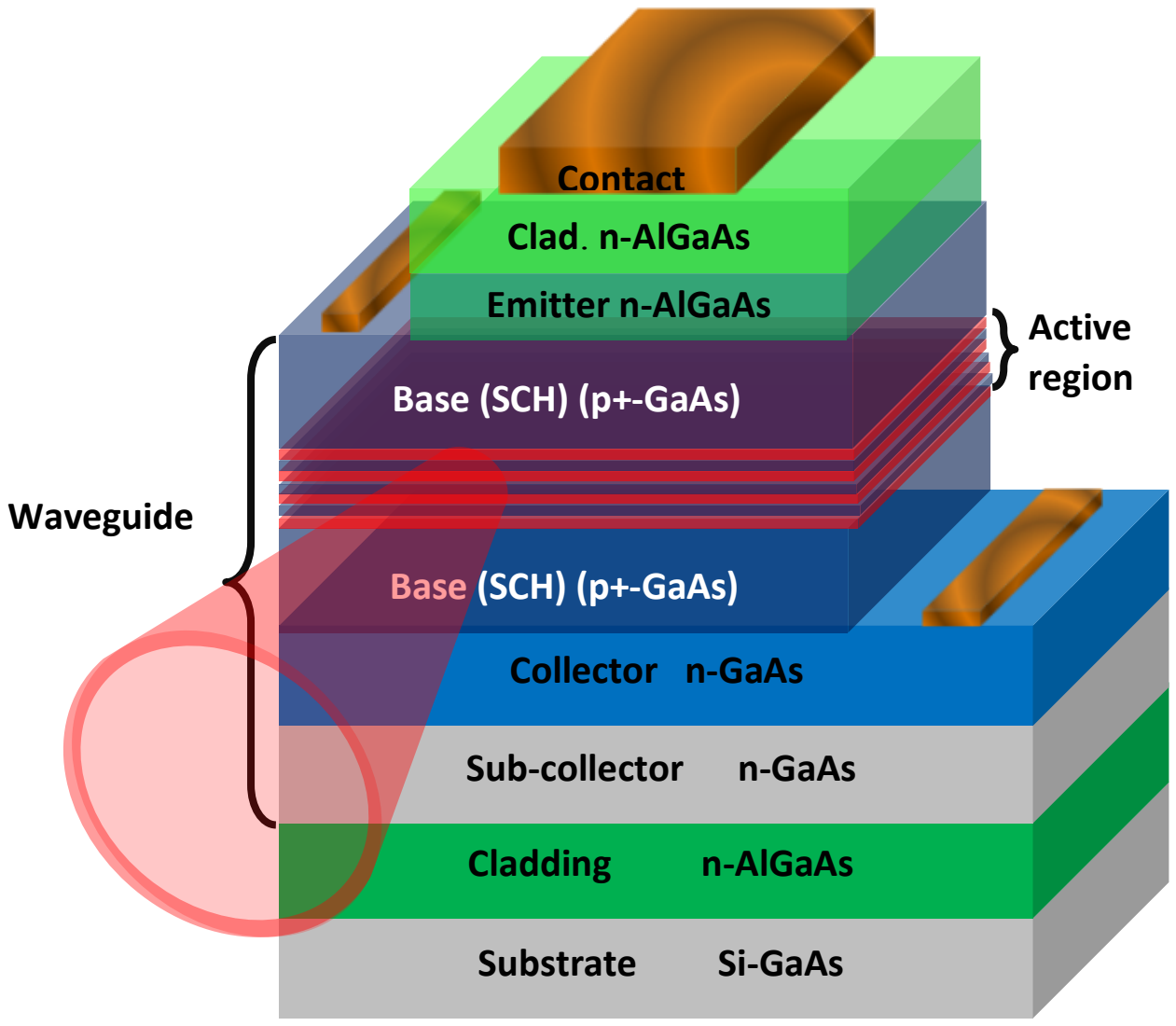

Fig. 1. Schematic illustration of an asymmetric MQW-HBTL. Emitter area, cavity length and facet reflectivity respectively are: $A = 1000\ \mu m^2, l = 200\ \mu m$ and $R_{1,2} = 0.32$

It is expectable that processing speed can be increased when the TL-equipped microprocessors are commercialized in the future.

HBTL has two new important features compared to the conventional DL. First, it is characterized by faster carrier recombination in the heavily doped base region, which improves the optical bandwidth. Second, resonance-free optical frequency response, as well as a higher level of injected carriers into QW, are realized by reversely biased collector-base junction in the HBTL [12]. In other words, the excess carriers are swept towards the electrical collector, i.e. collector-base interface, at a much higher speed than in conventional DLs. Optical and electrical characteristics of the HBTL have been studied both experimentally [10-12] and theoretically [13-17] to a significant extent. However, theoretical efforts to model HBTL's have been limited to reproducing specific electrical and optical operations of present devices in the small-signal regime, while a comprehensive study of large-signal behavior of the device is presently lacking. In [18], a large-signal analysis of TLs based on a charge control model was reported. In that work, the authors investigated properly the large-signal responses of TL for both sinusoidal and square wave modulation at different frequencies, but in their model and its large-signal analysis, only "one" quantum-well in the active region was assumed which is insufficient for the usual multi-quantum well TLs. On the other hand, a large-signal analysis of a buried hetero (BH) structure TL for an arbitrary number of quantum wells was also successfully investigated in [19]. However, assuming a uniform carrier distribution between QWs (we will show this assumption is incorrect), neglecting the tunneling effect between QWs and assigning fixed values for variable parameters (e.g., optical confinement factor, photon lifetime) which vary during the variation of the number of QWs, are the shortcoming of that work.

In this study, we show the large-signal analysis of a multi-QW transistor laser and the effect of carrier distribution and frequency on the line-width, using the device parameters extracted of structural factors [20]. Moreover, the turn-on time of a MQW-TL and its dependency on the structural factors is investigated.

## 2. MODEL

Transistor Laser studied here is based on N-p+-i heterojunction bipolar transistor (N-AlGaAs/p+-GaAs/i-GaAs) [21]. The TL operates in the active mode (base-emitter junction is forward-biased and base-collector is in reverse-biased condition). Demonstrated in Fig. 2 is the simulated band diagram of our assumed TL.

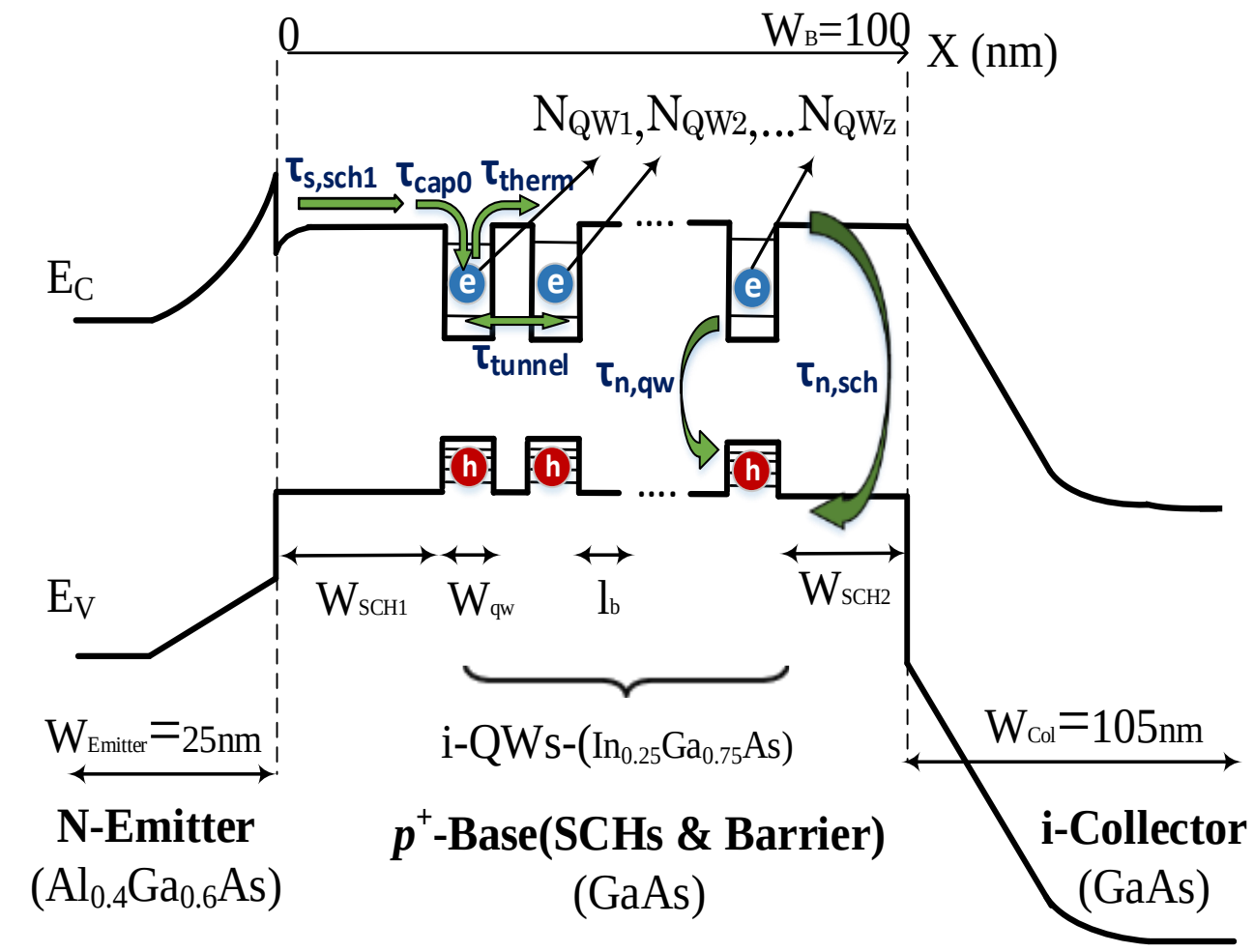

Fig. 2. Band diagram of Emitter/Base/Collector regions of a HBTL with intrinsic QWs. Electrons are diffuse across the $SCH_1$ by the carrier transport time $\tau_{s,SCH}$, captured into the quantum wells and then recombined by the QW lifetime $\tau_{n,qw}$. Carriers transport between QWs via tunneling ($\tau_{tunnel} = 0.824ps$) or a three-level mechanism of escaping ($\tau_{therm} = 400ps$), diffusing over the barrier and capture again into another QW by capture lifetime $\tau_{cap0}(= 0.22ps\ )$ [20]. The values of other parameters are: $l_b(\text{barrier width}) = 8nm$, $W_{qw}(QW\ width) = 7nm$ , $W_B(base\ width) = 100nm$.

In this paper, we take advantage of a previously-developed charge control model in [20] to simulate carrier transport within the base region including the active region. The model consisted of multi-level, coupled photon-electron rate equations shown below. In order to obtain large-signal analysis, we need to numerically solve Eqs. (1)-(7) using customized, Computationally-efficient techniques.

$$\frac{dN_{sch1}}{dt} = \frac{\eta_i J}{qW_{sch1}} - N_{sch1}\sum_{z=1}^{M}\frac{v_z}{\tau_{cap,z}} - \frac{N_{sch1}(1-\sum_{z=1}^{M} v_z)}{\tau_{n,sch1}} \quad (1)$$

$$\frac{dN_{QW1}}{dt} = N_{sch1}\frac{v_1}{\tau_{cap,1}}\frac{W_{sch1}}{W_{qw}} - \frac{N_{QW1}}{\tau_{therm}} - \frac{N_{QW1}}{\tau_{n,qw1}} - \frac{N_{QW1}-N_{QW2}}{\tau_c} - v_g g(N_1)(1-\varepsilon\Gamma S)S \quad (2)$$

$$\frac{dN_{QW_z}}{dt} = N_{sch1}\frac{v_z}{\tau_{cap,z}}\frac{W_{sch1}}{W_{qw}} + \frac{N_{QW_z}-N_{QW_{z-1}}}{\tau_c} - \frac{N_{QW_z}}{\tau_{therm}} - \frac{N_{QW_z}}{\tau_{n,qwz}} - \frac{N_{QW_z}-N_{QW_{z+1}}}{\tau_c} - v_g g\left(N_{QW_z}\right)(1-\varepsilon\Gamma S)S \quad (3)$$

$$\frac{\mathrm{d}N_{QW_M}}{\mathrm{d}t} = N_{sch1}\frac{\upsilon_M}{\tau_{cap,M}}\frac{W_{sch1}}{W_{qw}} + \frac{N_{QW_M} - N_{QW_{M-1}}}{\tau_c} - \frac{N_{QW_M}}{\tau_{therm}} - \frac{N_{QW_M}}{\tau_{n,qwz}} - \nu_g g\left(N_{QW_M}\right)(1-\varepsilon\Gamma S)S \tag{4}$$

$$\frac{\mathrm{dN}_{sch2}}{\mathrm{d}t} = \frac{N_{sch1}(1-\sum_{z=1}^{M}\upsilon_z)}{\tau_{n,sch2}} + \frac{N_{QW_M}}{\tau_{therm}}\frac{W_{qw}}{W_{sch2}} - \frac{N_{sch2}}{\tau_{n,sch2}}, \tag{5}$$

$$\frac{\mathrm{dS}}{\mathrm{d}t} = \Gamma\nu_g\left[\sum_{z=1}^{M} g\left(N_{QW_z}\right)\right] \times (1-\varepsilon\Gamma S)S - \frac{S}{\tau_p} + \frac{\beta}{M}\sum_{z=1}^{M}\frac{N_z}{\tau_{n,qwz}} \tag{6}$$

$$g\left(N_{\mathrm{QW}}\right) = G_0 \ln\left(\frac{AN_{QW} + BN_{QW}^2 + CN_{QW}^3}{AN_0 + BN_0^2 + CN_0^3}\right) \tag{7}$$

where $z$ is represented for the QW number which varies from $z = 0$ to M (number of QWs), $\mathrm{N}_{sch}$ and $W_{sch}$ are the carrier density and the width of SCHs region respectively, $N_{QW_z}$ and $g\left(N_{QW_z}\right)$ are the $z^{th}$ quantum well carrier density and carrier-dependent optical gain, $\eta_i(\approx 1)$ is internal quantum efficiency, $\beta\ (= 1 \times 10^{-4})$ is the spontaneous emission factor, $J$ is the base current density, $q$ is the electron charge, $\tau_{n,sch}$ and $\tau_{n,qw}$ are the carrier recombination lifetime in SCH region and QWs respectively, Γ is the optical confinement factor (variable for different number of QWs) [20], $\varepsilon\ (= 1.9 \times 10^{-17}\ cm^3)$ is the gain compression factor, $v_g(= 8.7 \times 10^9\ cm.s^{-1})$ is the group velocity, S is the photon density, $\tau_p$ is the photon lifetime, $\tau_c$ is the carrier inter-well transport time [20], $\tau_{cap,z} = (W_{\mathrm{sch}}^2/2D) + ((z-1)\,l_b^2/2D) + \tau_{cap_0}$ is the overall capture lifetime for $z^{th}$ QW, $D(= 28.65\ cm^2s^{-1})$ is the diffusion constant [22, 23]. $G_0\ (= 1370\ cm^{-1})$ is the gain coefficient [24], $N_0\ (= 1.26 \times 10^{18}\ cm^{-3})$ is the transparency carrier density and $A, B$ and $C$ are the monomolecular, bimolecular and Auger recombination coefficients respectively which are extracted from experimental data [25, 26]. Moreover, $\upsilon_z$ is the $z^{th}$ QW geometry factor which is defined as the fraction of base charges captured by the $z^{th}$ well [5].

More precise optical gain associated with the $z^{th}$ quantum well which we use it further to show its variation with wavelength is formalized by:

$$G_z\left(N_{QW_z}, P_{QW_z}, S\right) = \Gamma\nu_g G\left(N_{QW_z}, P_{QW_z}, S, \omega\right), \tag{8}$$

where $v_g$ is the group velocity and Γ is the optical confinement factor of the active region. In a separate study, we have presented a calculation technique for Γ in the MQWHBTL as a function of the device structural factors [20] that is used for the calculation of $G_z$.

In addition, the material gain can be obtained as:

$$G_z\left(N_{QW_z}, P_{QW_z}, S, \omega\right) = \frac{\omega}{c_0\varepsilon_0\bar{n}^2}\sum_{n=1}^{\psi}\int_{E_{c-hh}^n}^{\infty} \rho|\mu|^2 \frac{(f_c - f_{hh})\pi L}{1 + S/S_{sat}} \tag{9}$$

$$f_c = \left[1 + \exp\left(\frac{E_e^n - F_c}{k_B T}\right)\right]^{-1} \tag{10}$$

$$f_{hh} = \left[1 + \exp\left(\frac{E_{hh}^n - F_{hh}}{k_B T}\right)\right]^{-1} \tag{11}$$

where $N_{qw}$ and $P_{qw}$ are the electron and hole densities inside the QW, S is the photon density, $S_{sat}$ is the saturation photon density, ω is the angular frequency, $c_0$ is the light speed, $\varepsilon_0$ is the dielectric constant, $f_c$ and $f_{hh}$ are the electron and heavy holes Fermi distributions, $k_B$ is the Boltzmann constant, $T$ is temperature, $E_e^n$ and $E_{hh}^n$ are the energy states of electrons and heavy holes respectively, $F_c$ and $F_{hh}$ are the carrier-dependent quasi-Fermi levels of the conduction and valence band, L is the Lorentzian lineshape function, $\rho$ is the density of dipole states, μ is the dipole moment for the TE mode, respectively. We leave the details of the optical gain calculation to readers as separately described in a separate work [24].

3. Large-signal analysis

Considering the highly doped base region (hole density>$10^{19}$), we assumed a quasi-equal condition in which $P_{qw} \approx N_{qw}$. Simulation results for optical gain are fed into a multi-level, charge control model describing the carrier-photon rate equations. For large signal analysis, we need to develop an iterative computational method for solving charge control and gain model at the same time. In order to better illustrate of second-order phenomena of large-signal analysis, we performed our numerical solving for two different (low and high) resolutions. In Fig. 3, we can see good accordance between small-signal and large-signal modulation response for two resolutions of a typical single-QW HBTL.

Our studies showed that the photon density in this regime has a prominent effect on the optical bandwidth of the device. Fig. 4 shows the "low resolution" simulation results for such a study in which we considered both single and double QW structures in response to a large signal expressed by $J(t) = J_0 + \Delta J_m \times \sin(\omega t)$ in which $J_0$ and $\Delta J_m$ are the current density at DC and AC terms, ω is the angular frequency and J(t) is the time-dependent current density feeding the TL. (7) As can be seen in Fig. 4 (a), 2QW-HBTL generates higher photon density due to amplified "optical" behavior compared to the SQW case. This agrees with the gain-bandwidth trade-off expressed in [27] and [28].

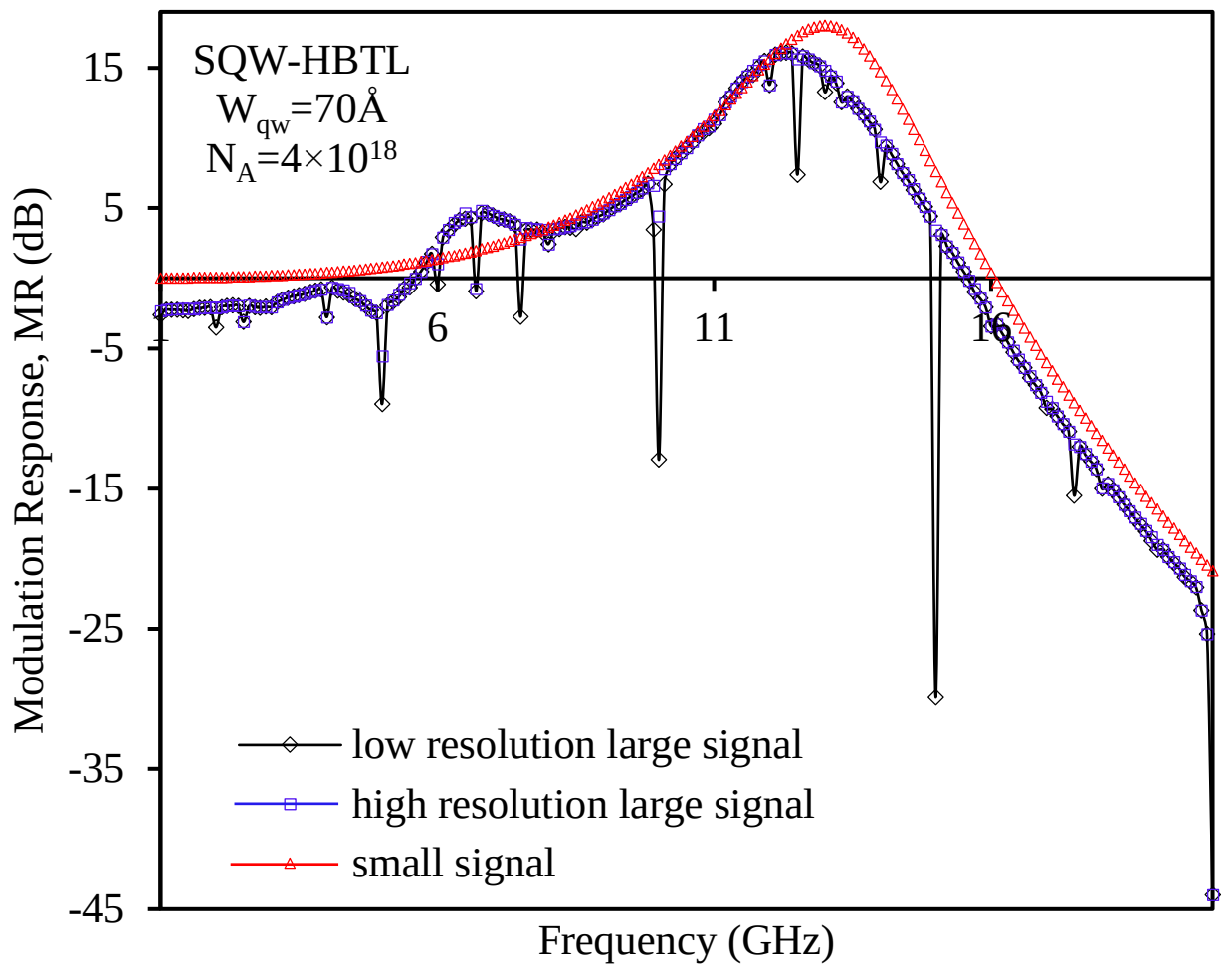

Fig.3. Modulation responses for both large and small-signal regimes of a typical SQW-TL with QW width and base doping of $W_{qw} = 7nm$ and $N_A = 4 \times 10^{18}\ cm^{-3}$ respectively. The large-signal analysis is done for two different resolutions.

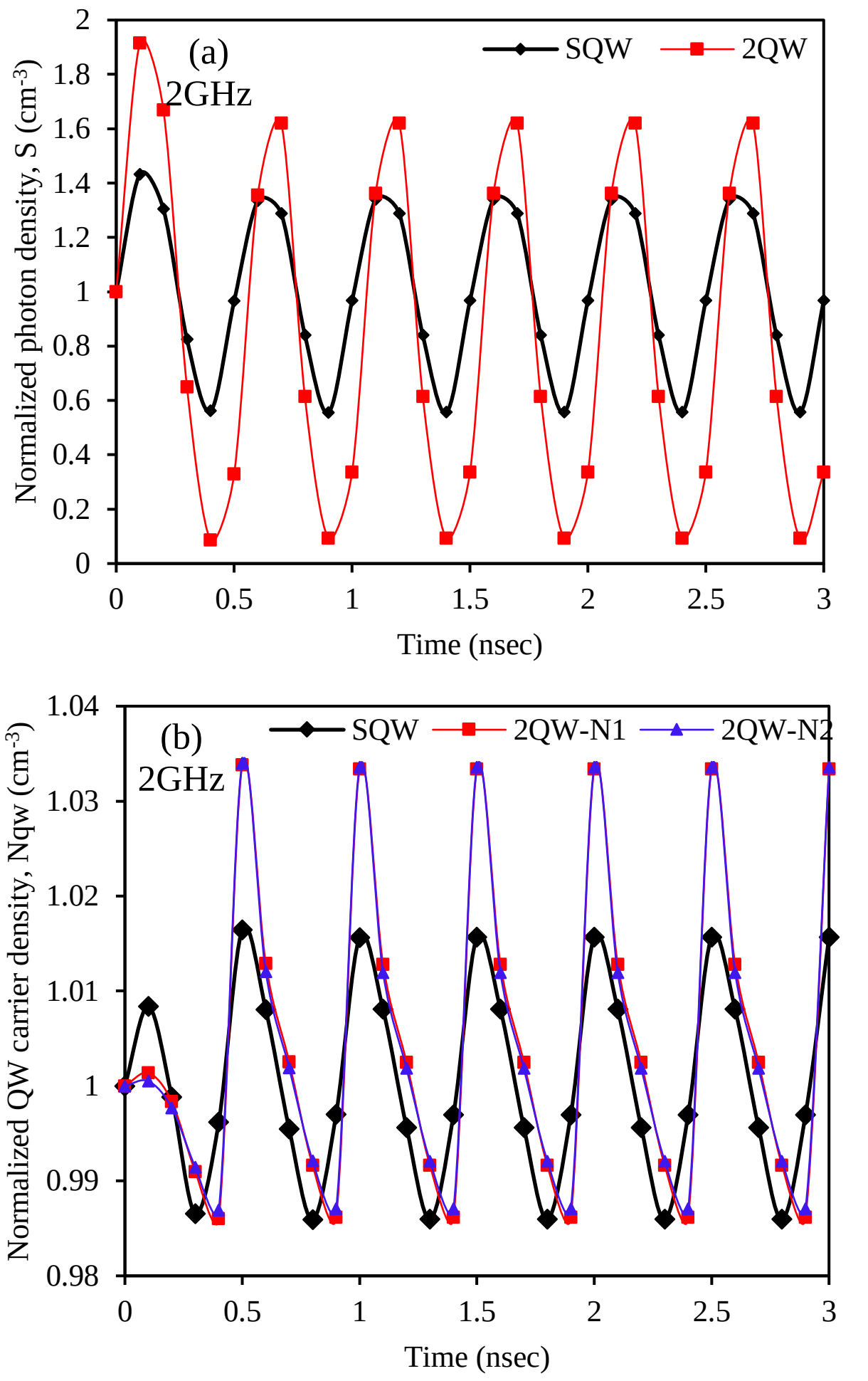

Fig. 4. Comparison between SQW and 2QW to a low resolution large signal; J(t)= $J_0$+ ΔJm×sin(ωt) (J0=6000 A/cm2, ΔJ= J0/3) (a) photon density response (b) carrier density response.

Specifically, we notice an increased photon density as a result of adding a QW. This should reduce current gain (β) to 50% of its value. The carrier density shown in Fig. 4 (b) is simulated at low enough frequencies. It approves there is no significant difference between the first QW and the second QW. We expect, however, the difference should slowly rise with increasing the number of QWs.

An interesting large-signal aspect of TL shows itself in the "high resolution" numerical computing. As can be seen in Fig. 5, two nodes at every half-period of photon densities can be observed. Although MQW-HBTLs have higher bandwidth and optical output power than SQW-HBTLs, this phenomenon can be a shortcoming of MQW-TLs due to increasing the BER (Bit Error Rate).

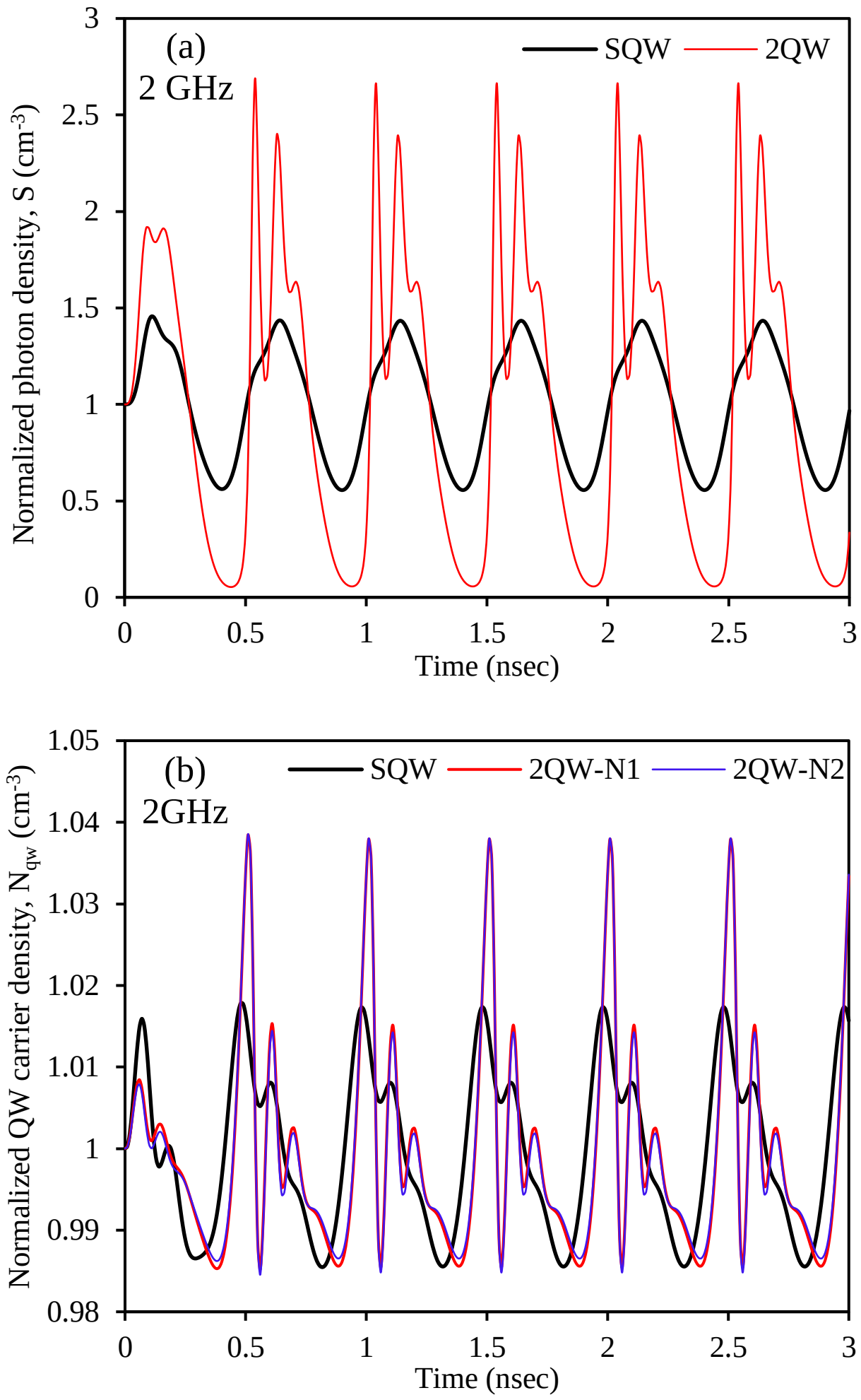

Fig. 5. High resolution responses of photon (a) and carrier (b) densities to a large-signal sinusoidal modulation of base current (J(t)= $J_0$+ ΔJm×sin(ωt); J0=6000 A/cm2, ΔJ= J0/3) for SQW and 2QW structures.

Modeling the large-signal behavior of the device to higher frequencies reveals that we have a noticeable imbalanced carrier distribution between subsequent QWs, which increases with frequency. Fig. 6 shows the large signal response of the proposed TL with a 3QW structure to an electrical input at 4 GHz, 12GHz and 18 GHz. Interestingly, as frequency goes up, the difference between carrier densities in different quantum wells gradually increases and a decrease of up to ~26% is observed in carrier density of the last QW compared to the first one ($\Delta N_{qw}=N_{qw,1}-N_{qw,n}$) at 18 GHz.

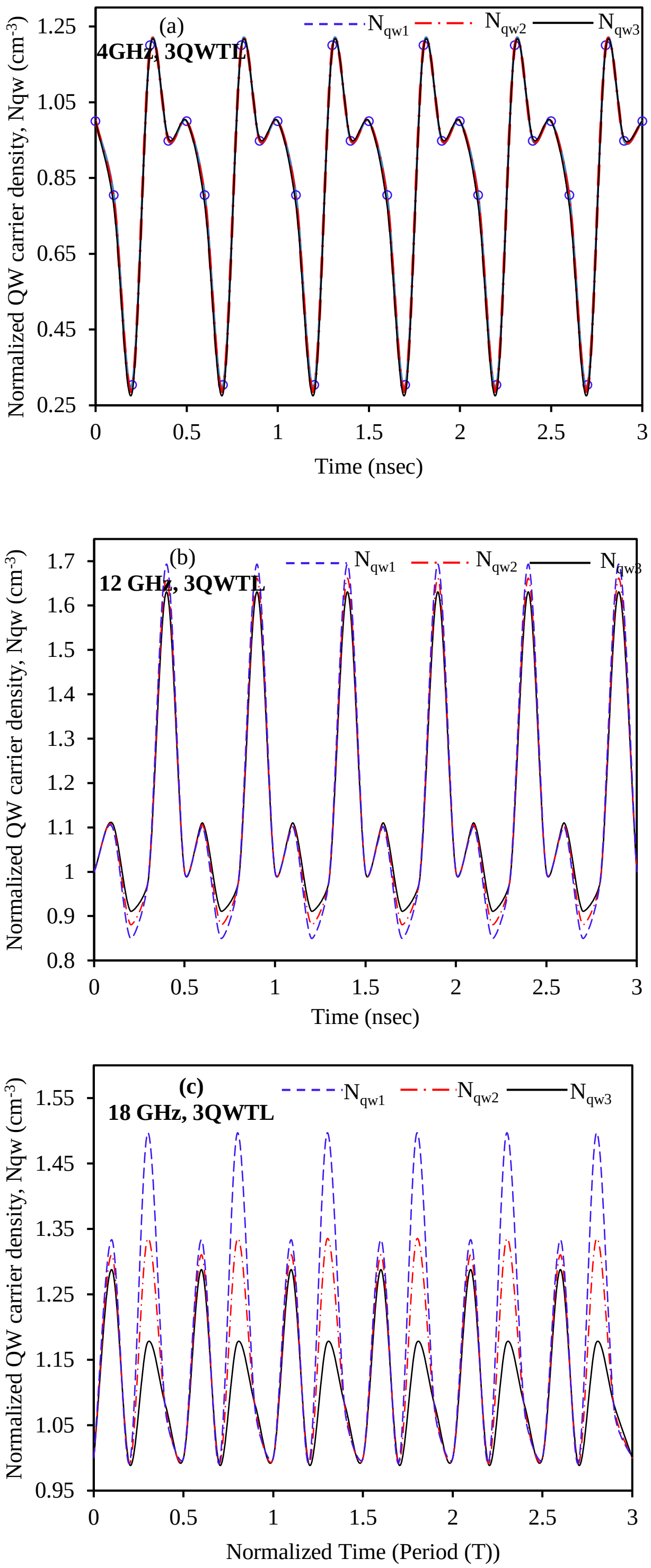

Fig. 6. Imbalanced carrier density distribution at 4 GHz (a), 12 GHz (b) and 18 GHz (c) large signal. Note: Time (x-axis) is scaled and normalized to the period of each individual period.

It is obvious that the barrier width can play a noticeable role in minimizing this effect. Following Eqs. (8)-(11), the material gain associated with each QW is proportional with $N_{qw}$, or equivalently $\Delta N_{qw}$, hence the total optical gain of the device will demonstrate a larger linewidth.

Material gain calculation results for two typical transistor lasers, i.e. single QW (SQW) and three QW (3QW), are shown in Fig. 7. $\Delta\lambda_{0\text{-}0}$ and $\Delta\lambda_{1\text{-}1}$ in Fig. 7(a) respectively represent for the error in simulated peak wavelength of ground state and first excited state compared to experimental data [11]. For both SQW and 3QW structures we observed acceptable error in peak wavelength of the lasing inside the QW region [29]. This approves the accuracy of our gain calculation model to be employed later. The discussed phenomena of imbalanced carrier distribution within the QWs at high enough frequencies ($\Delta N_{qw}$) is a disadvantage for the MQW structure versus SQW that should be taken into consideration when laser linewidth is as important as the bandwidth.

.

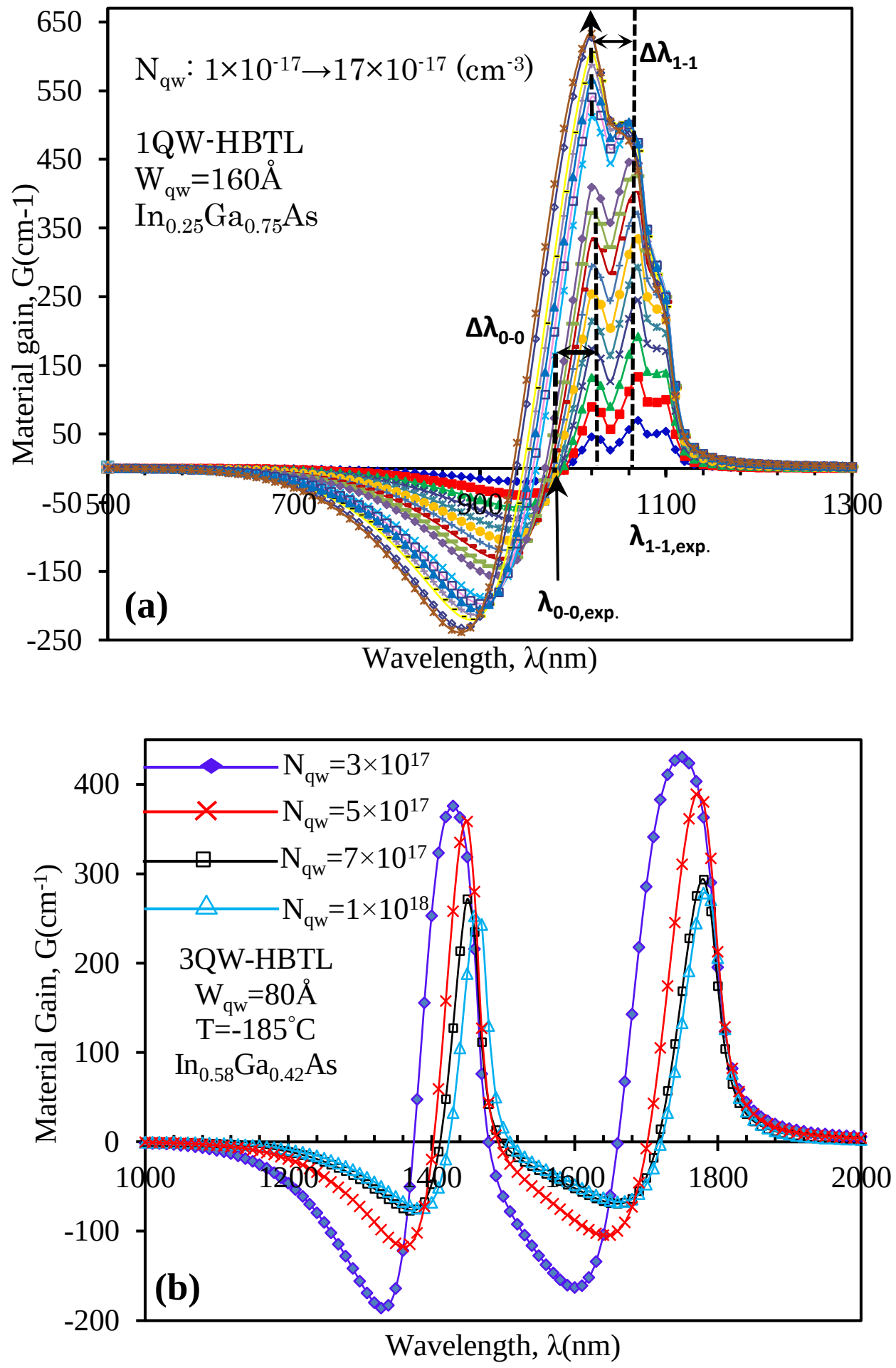

Fig. 7. Simulation results of material gain versus wavelength for (a) SQW-HBTL ($W_{qw}$=160Å, $In_{0.25}Ga_{0.75}As$). (b) 3QW-HBTL. It can be seen that the material gain is related to the QW carrier densities. Hence, the difference between QWs carrier densities at high frequencies causes a line-shape broadening.

4. Switching analysis

To obtain the turn-on characteristics of MQW-HBTL, the Eqs. (1)-(7) are numerically solved with efficient numerical methods. The turn-on time is defined as the time duration from when the base current pulse turns-on to when the stimulated emission occurs and the photon density starts to raise. Two important factors have an effect on turn-on time, i.e., threshold current level and the carrier redistribution-time (the time it takes by the carriers to redistribute between QWs via tunneling or other mechanisms). Fig. 8, shows the dependency of the turn-on time of MQW-HBTL on the injected current. By increasing the base current level, the differential gain of HBTL increases and it causes faster carrier distribution between QWs and of course, decreasing the turn-on time. On the other hand, as the number of QWs increases, the threshold current starts to decrease [20] and the redistribution time increases. This trade-off between threshold current level and redistribution time is obvious in Fig. 8 (inset).

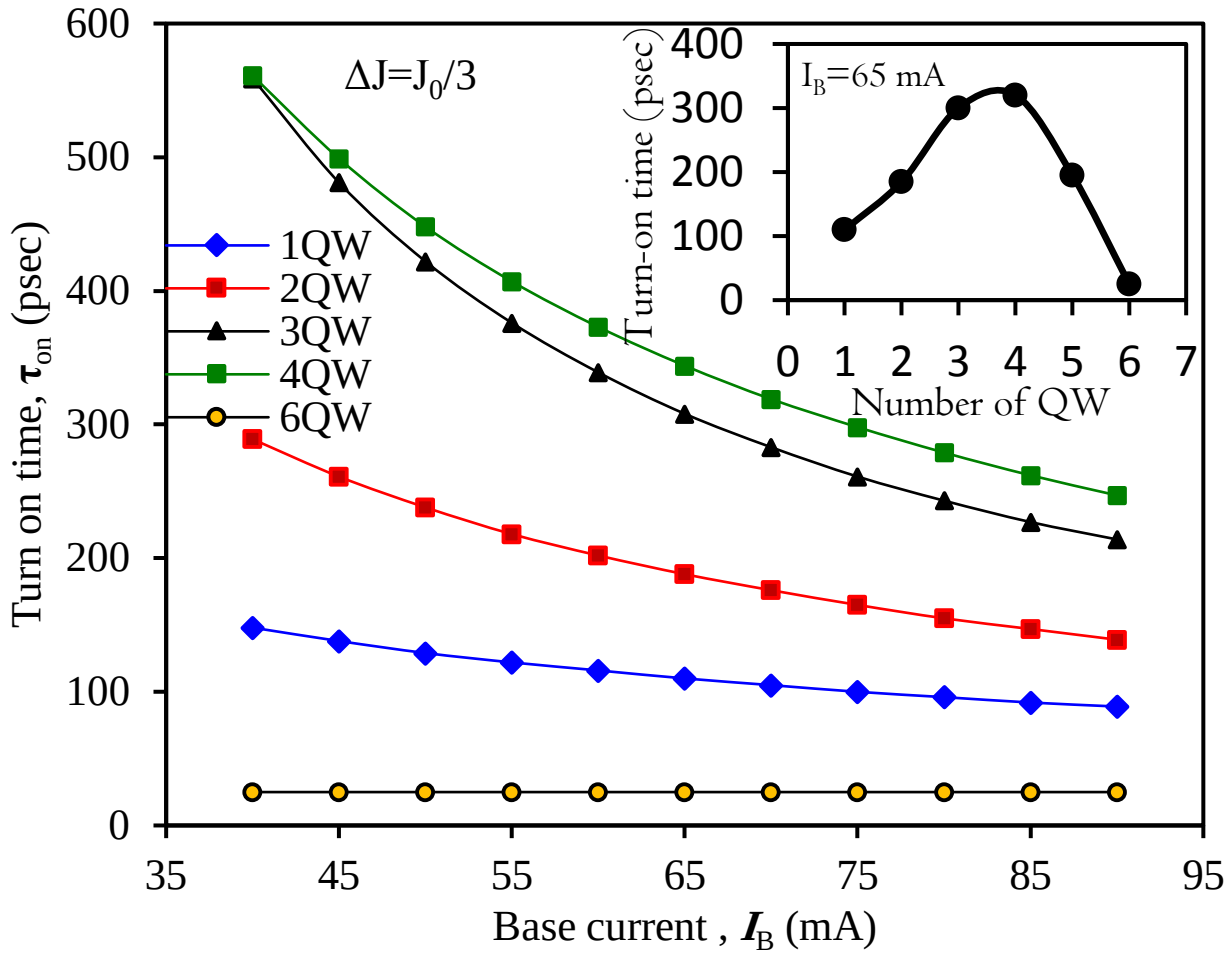

Fig. 8. The dependency of turn-on time of MQW-HBTL on the base current for different number of QWs. Turn-on time starts to decrease due to increasing the level of current injection. Inset: turn-on time versus the number of QW for a typical base current ($I_B$=65 mA).

An important parameter that affects the turn-on time by changing the redistribution time, is the barrier width. This factor shows itself in the inter-well carrier transport time, which comprise of two mechanisms i.e. 1) tunneling and 2) a three-step mechanism of escape, diffuse over the barrier and then capture in the QW which is defined as $\tau_c = 1/[1/\tau_{tunnel} + 1/(\tau_{therm} + \tau_{bar} + \tau_{cap,0})]$ where $\tau_{bar} = l_b^2/2D$ is the time a carrier takes to diffuse over the barrier. As the barrier width starting to decrease (lower than $l_b \approx 135A^\circ$ ), the dominant mechanism will be tunneling. In Fig. 9, we can see the sensitivity of the turn-on time of HBTL with 2-QW and 6-QW to the barrier width in comparison with SQW-HBTL.

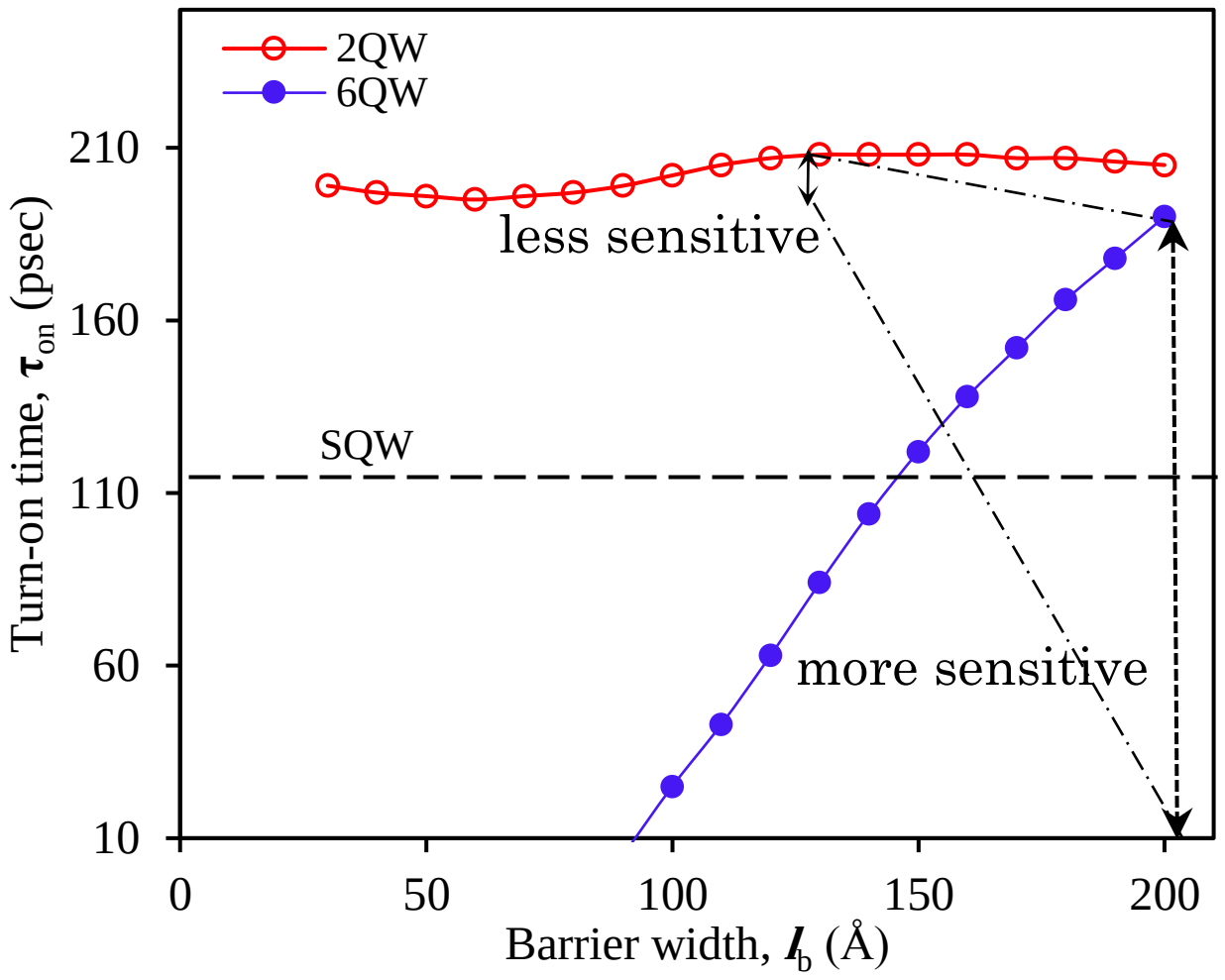

Fig. 9. Calculated turn-on time of single-QW, 2QW, and 6QW HBTL as a function of barrier width.

It can be understood from the Fig.9, that: 1) the turn-on time decreases as the barrier width decreases due to reduction of tunneling lifetime ($\tau_{tunnel}$) [19] and 2) the sensitivity of turn-on time to the barrier width increases as the number of QWs increases. So it can be possible to reach even lower than 100ps for turn-on time by utilizing six QWs in the active region and also narrowing the barrier width to its permitted values. Moreover, the turn-on time of a MQW-HBTL as a function of cavity length depicted in Fig. 10. It is expectable that turn-on time reduces by elongating the cavity length due to the reduction of cavity loss [30].

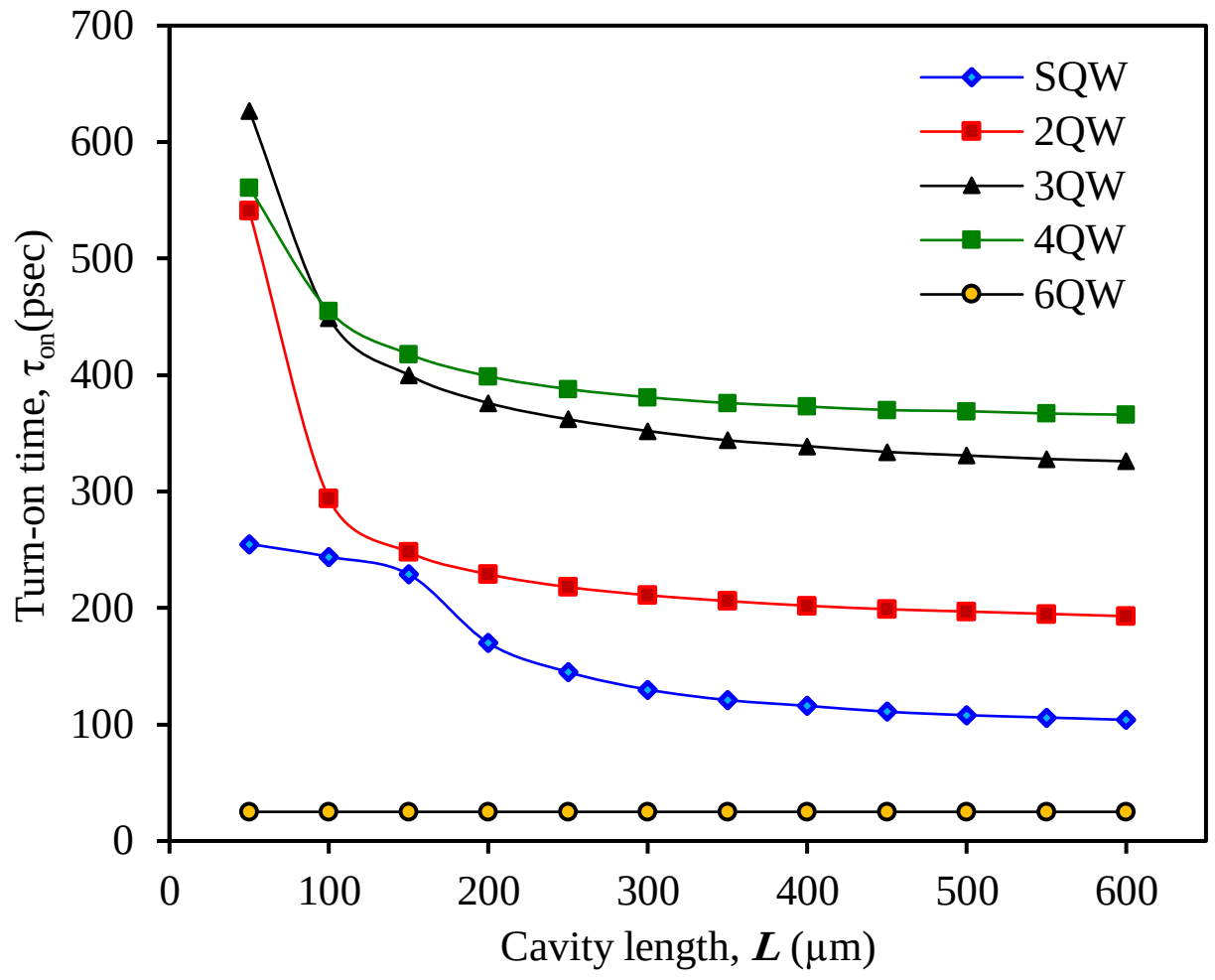

Fig. 10. Turn-on time versus cavity length for SQW and MQW transistor laser.

To utilize this interesting optoelectronic device in an optoelectronic integrated circuit (OEIC), we need a comprehensive optimization including dc, ac and switching performances to make a good judgment of this device. Accordingly, we define a parameter named "relative performance factor" for MQW-HBTL as:

$$\Theta = \frac{f_{-3dB}^{\theta_1} \times P_{out}^{\theta_2} \times \beta_{ac}^{\theta_3}}{I_{th}^{\theta_4} \times \tau_{on}^{\theta_5}}, \quad (12)$$

where $f_{-3dB}$, $P_{out}$, $\beta_{ac}$, and $I_{th}$ are the modulation bandwidth, output power, ac current gain and base threshold current respectively which completely discussed and calculated in [5, 20] and in order to magnify the importance of any of these parameters based on the application of TL in the Opto-Electronic Integrated Circuits (OEICs), five importance factor ($\theta_1$ to $\theta_5$) are taken into consideration. In this study, to overall optimizing, we assign equal significance to these importance factors and set them all to one ($\theta_1 = \theta_2 = \cdots = \theta_5 = 1$). In Fig. 11, the relative performance factor illustrated as a function of injected current and number of QWs. As can be seen, at low injected current density the optimum structure is SQW structure. However, at a higher level of current injection, the optimum structure transforms into structures with a higher number of QWs. As depicted in Fig. 11, for current density higher than $2000\ A.cm^{-2}$ and$4000\ A.cm^{-2}$, ($J_B > 2000\ A.cm^{-2}$ and $J_B > 4000\ A.cm^{-2}$), the optimum structures are 2QW and 3QW respectively. In other words, to fully exploit the MQW-HBTL capability, we need to use and bias them at a higher current density level.

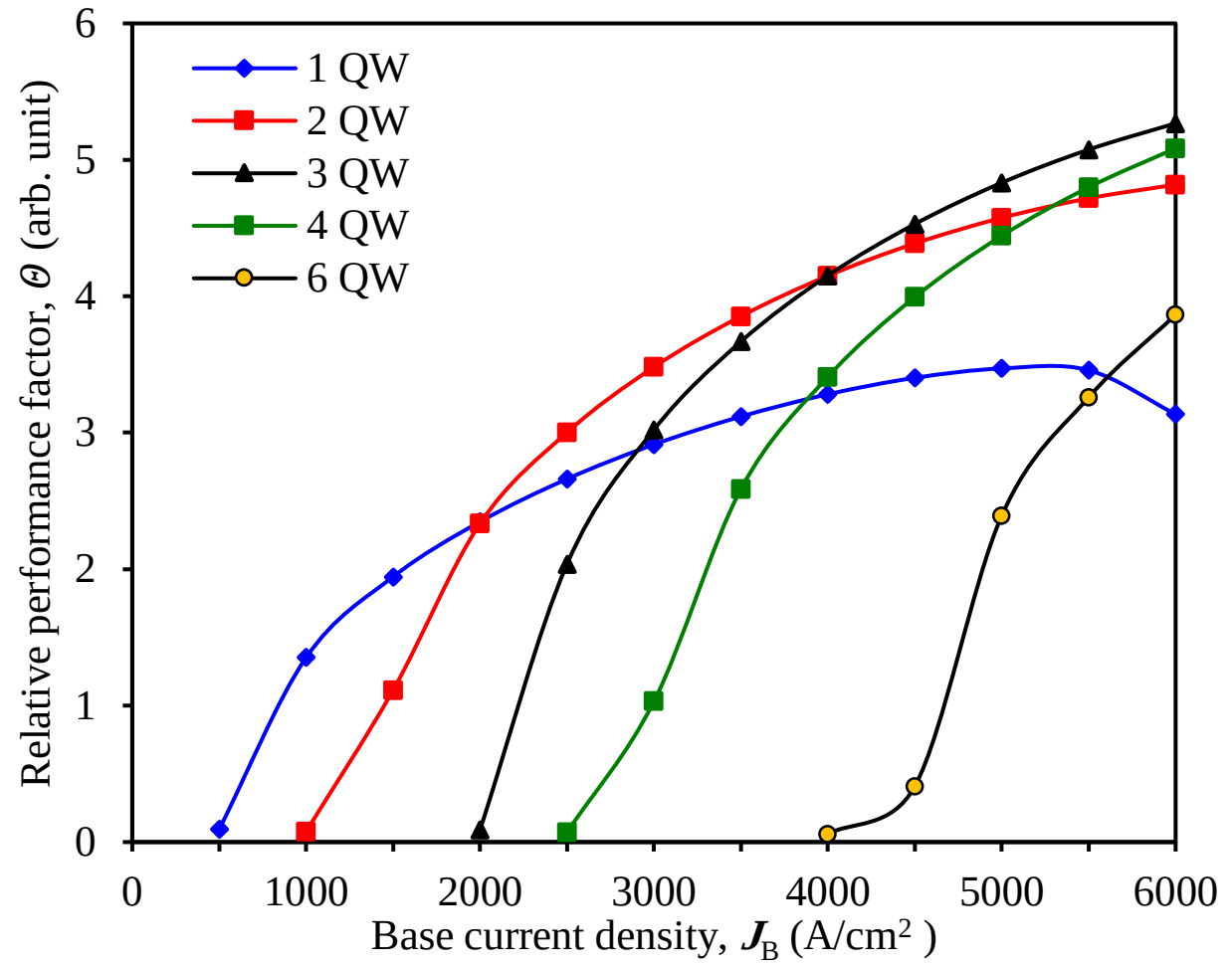

Fig. 11. Relative performance factor for MQW-HBTL as a function of injected current density. MQW-HBTL has a higher performance factor at higher base current density.

## 5. Conclusion

In conclusion, we numerically solved a set of coupled rate equations with a nonlinear carrier and photon-dependent optical gain to investigate the large-signal and switching behavior of a MQW-HBTL. Our large-signal analysis revealed a considerable imbalance carrier distribution between QWs at high frequencies which causes broadening line-shape. Turn-on times as a function of structural factors and bias condition (e.g. number of QWs, cavity length, barrier width, and base current density) were investigated. In order to minimize the turn-on time we can enlarge the cavity length or utilizing the HBTL with a higher number of QWs but thinner barrier width and also bias the HBTL at higher base current. We also introduced a relative performance factor to anticipate the optimum structure for MQW-HBTL. In other words, not only the dc and ac characteristics such as the modulation bandwidth, output power, ac current gain, and base threshold current but also switching characteristics such as turn-on time should be taken into account for a comprehensive optimization. Based on the relative performance factor, the more QW we add to the active region, the more performance we can obtain of TL, but at a higher level of base current and of course higher power consumption.